\documentclass[]{piparticle-final}
\usepackage{graphicx}
\usepackage{amsmath}
\begin{document}

\volume{2}               
\articlenumber{020002}   
\journalyear{2010}       
\editor{D. A. Stariolo}   
\received{23 December 2009}     
\accepted{24 February 2010}   
\runningauthor{G. J. Zarragoicoechea \textit{et al.}}  
\doi{020002}         

\title{Multilayer approximation for a confined fluid in a slit pore}


\author{G. J. Zarragoicoechea,\cite{inst1,inst2}%
\thanks{E-mail: vasco@iflysib.unlp.edu.ar}\hspace{0.5em}  A. G. Meyra,\cite{inst1} V. A. Kuz\cite{inst1}}

\pipabstract{A simple Lennard--Jones fluid confined in a slit nanopore
with hard walls is studied on the basis of a multilayer structured
model. Each layer is homogeneous and parallel to the walls of the
pore. The Helmholtz energy of this system is constructed following
van der Waals-like approximations, with the advantage that the model
geometry permits to obtain analytical expressions for the integrals involved. Being the multilayer system in thermodynamic equilibrium,
a system of non-linear equations is obtained for the densities and
widths of the layers. A numerical solution of the equations gives the
density profile and the longitudinal pressures. The results are compared
with Monte Carlo simulations and with experimental data for Nitrogen,
showing very good agreement.}

\maketitle

\blfootnote{ \begin{theaffiliation}{99} \institution{inst1}
IFLYSIB-Instituto de F\'{\i}sica de L\'{\i}quidos y Sistemas Biol\'{o}gicos
(CONICET, UNLP, CICPBA), 59 No. 789, 1900 La Plata, Argentina. \institution{inst2}
CICPBA-Comisi\'{o}n de Investigaciones Cient\'{\i}ficas de la Prov. de
Buenos Aires. \end{theaffiliation} }

\section{Introduction}

The effects on phase transition of confined fluids in a slit-like
pore have been studied by simulation and different theories {[}1--11{]}.
In a previous work, we constructed a generalized van der Waals equation
for a fluid confined in a nanopore {[}12, 13{]}. The shift of the
critical parameters was in good agreement with lattice model and numerical
simulation results, and the predicted critical temperature remarkably
reproduced the experiment. In that work, we concluded that the confined
van der Waals fluid theory seemed to work better than the bulk one,
maybe due to the fact that the higher virial contributions not considered
in both theories were less important in the confined fluid than in
the bulk. A similar treatment was used previously by Schoen and Diestler
{[}14{]}. Following that line of reasoning, here we study a simple
fluid confined between two infinite parallel hard walls (slit pore).
The walls are at a distance $\mathit{L}$ apart. To study the confined
fluid, we propose a multilayer model {[}15{]}: the fluid is distributed
in \textit{n} thin layers, one beside the other. Each layer has a
uniform density, and can be observed as a non-autonomous phase. A
particle in a given layer interacts with its neighbors inside the
layer, and with every particle in the other layers. Defay and Prigogine
and Murakami et al. have shown that, in a liquid gas interface, the
deviation from the Gibbs' adsorption equation becomes practically
negligible in the case of a two layer model {[}16{]}, and that
as the number of transition layers grows, the multilayer model becomes
perfectly consistent with the Gibbs' equation {[}17{]}. The van der
Waals-like approximations made in developing this multilayer model
theory limit its validity to the low density regime.

\section{Theory}

The model system consists of a fluid of $\mathit{N}$ Lennard--Jones
particles confined in a slit nanopore. The hard walls of the pore,
separated at a distance $\mathit{L}$ (in the $\mathit{x}$ direction),
have a surface area $\mathit{S}$ ($\mathit{S}\rightarrow\infty$).
We divided the fluid into $\mathit{n}$ layers, each layer being
parallel to the pore walls. The layer $\mathit{i}$\textit{\ } has
$\mathit{N}_{\mathit{i}}$\textit{\ }particles ($\mathit{N}=\sum_{i=1}^{n}N_{i}$),
a width $\mathit{L}_{\mathit{xi}}$ ($\mathit{L}$= $\sum_{i=1}^{n}L_{xi}$),
and a volume $V_{i}=SL_{xi}$. Then the Helmholtz energy {[}18{]}
can be written as

\begin{equation}
A=-kT\ln\left(\frac{Z_{N}\lambda^{-3N}}{\prod\limits _{i=1}^{n}N_{i}!}\right)\label{GrindEQ__1_}.\end{equation}
The configuration integral $\mathit{Z}_{\mathit{N}}$ for a pair potential
$\mathit{v}_{\mathit{ij}}$ may be approximated as

\begin{eqnarray}
Z_{N} & = & \int\prod\limits _{\substack{i=1\\
i<j}
}^{n}e^{-v_{ij}/kT}\mathrm{d}\mathbf{r}^{N}\approx\sum_{\substack{i=1\\
i<j}
}^{n}\int f_{ij}\mathrm{d}\mathbf{r}^{N}\label{GrindEQ__2_}\\
 &  & +\prod\limits _{i=1}^{n}V_{i}^{N_{i}}\notag\\
f_{ij} & = & e^{-v_{ij}/kT}-1,\notag\end{eqnarray}
and further expanded in function of two particle integrals

\begin{eqnarray}
{Z_{N}} & {=} & {\sum\limits _{i=1}^{n}\frac{N_{i}(N_{i}-1)}{2}V_{i}^{N_{i}-2}}\label{GrindEQ__3_}\\
 &  & {\prod\limits _{\substack{k=1\\
k\neq i}
}^{n}V_{k}^{N_{k}}\int\limits _{R_{i}}f_{12}\mathrm{d}}\mathbf{r}{_{\mathbf{1}}\mathrm{d}\mathbf{r}_{\mathbf{2}}}\notag\\
 &  & +{\sum\limits _{i=1}^{n}\sum\limits _{\substack{j=2\\
j>i}
}^{n}N_{i}N_{j}V_{i}^{N_{i}-1}V_{j}^{N_{j}-1}}\notag\\
 &  & {\prod\limits _{\substack{k=1\\
k\neq i,k\neq j}
}^{n}V_{k}^{N_{k}}\int\limits _{R_{i}}\int\limits _{R_{j}}f_{12}\mathrm{d}}\mathbf{r}{_{\mathbf{1}}\mathrm{d}}\mathbf{r}{_{\mathbf{2}}+\prod\limits _{i=1}^{n}V_{i}^{N_{i}}}.\notag\end{eqnarray}
The first term in Eq. (\ref{GrindEQ__3_}) stands for particles in
the layer $\mathit{i}$. The second term comes from the interaction
of one particle in layer $\mathit{i}$ with one particle in layer
$\mathit{j}$\textit{.} In a compact form, and assuming that a layer
sees three nearest neighbor layers,

\begin{eqnarray}
Z_{N} & = & (\sum\limits _{i=1}^{n}\frac{N_{i}^{2}}{2V_{i}^{2}}I_{i}\label{GrindEQ__4_}\\
 &  & +\sum\limits _{i=1}^{n-1}\sum\limits _{\substack{j=i+1\\
j\leq n}
}^{i+3}\frac{N_{i}}{V_{i}}\frac{N_{j}}{V_{j}}I_{i\, j}+1\,)\prod\limits _{i=1}^{n}V_{i}{}^{N_{i}}.\notag\end{eqnarray}
The integrals $\mathit{I}_{\mathit{i}}$ and $\mathit{I}_{\mathit{ij}}$,
for the slit pore geometry and after low density approximations, can
be analytically solved to give

\begin{equation}
\begin{array}{l}
{I_{i}=\iint\limits _{R_{i}}f_{12}\mathrm{d}\mathbf{r}_{1}\mathrm{d}\mathbf{r}_{2}\approx-\iint\limits _{\left\vert \mathbf{r}_{1}\mathrm{-}\mathbf{r}_{2}\right\vert <\sigma}\mathrm{d}\mathbf{r}_{1}\mathrm{d}\mathbf{r}_{2}}\\
{-\iint\limits _{\left\vert \mathbf{r}_{1}\mathrm{-}\mathbf{r}_{2}\right\vert \geq\sigma}\frac{v_{12}}{kT}\mathrm{d}\mathbf{r}_{1}\mathrm{d}\mathbf{r}_{2}=-2V_{i}\sigma^{3}(b-B_{i})-\frac{2V_{i}\sigma^{3}\varepsilon}{kT}A_{i}}\end{array}\label{GrindEQ__5_}\end{equation}

\begin{equation}
\begin{array}{l}
{I_{i,\, i+1}=\int\limits _{R_{i}}\int\limits _{R_{i+1}}f_{12}\mathrm{d}\mathbf{r}_{1}\mathrm{d}\mathbf{r}_{2}\approx-\iint\limits _{\left\vert \mathbf{r}_{1}\mathrm{-}\mathbf{r}_{2}\right\vert <\sigma}\mathrm{d}\mathbf{r}_{1}\mathrm{d}\mathbf{r}_{2}}\\
{-\iint\limits _{\left\vert \mathbf{r}_{1}\mathrm{-}\mathbf{r}_{2}\right\vert \geq\sigma}\frac{v_{12}}{kT}\mathrm{d}\mathbf{r}_{1}\mathrm{d}\mathbf{r}_{2}=-V_{i}\sigma^{3}B_{i}-\frac{V_{i}\sigma^{3}\varepsilon}{kT}A_{i,\, i+1}}\end{array}\label{GrindEQ__6_}\end{equation}

\begin{eqnarray}
\begin{array}{l}
I_{i,\, i+2} = \int\limits _{R_{i}}\int\limits _{R_{i+2}}f_{12}\mathrm{d}\mathbf{r}_{1}\mathrm{d}\mathbf{r}_{2}\approx\\
-\iint\limits _{\left\vert \mathbf{r}_{1}\mathrm{-}\mathbf{r}_{2}\right\vert \geq\sigma}\frac{v_{12}}{kT}\mathrm{d}\mathbf{r}_{1}\mathrm{d}\mathbf{r}_{2} = -\frac{V_{i}\sigma^{3}\varepsilon}{kT}A_{i,\, i+2}\end{array}\label{GrindEQ__7_}\end{eqnarray}

\begin{equation}
I_{i,\, i+3}=-\frac{V_{i}\sigma^{3}\varepsilon}{kT}A_{i,\, i+3}\label{GrindEQ__8_}\end{equation}
In the above expressions $\mathit{v}_{\mathit{ij}}$ was taken to
be the Lennard--Jones pair interaction, being $\varepsilon$\ and
$\sigma$\ the potential parameters. The integrals $I_{i,\, i+2}$
and $I_{i,\, i+3}$\ do not contain the excluded volume term because
we suppose that the layer widths are $\mathit{L}_{\mathit{xi}}\geq\sigma$.
The expressions for $\mathit{A}$ and $\mathit{B}$ in the preceding
equations, functions of $\mathit{L}_{\mathit{xi}}$, are given in
the Appendix A.

The Helmholtz energy, Eq. (1) together with Eq. (4), has the final expression

\begin{equation}
\begin{array}{l}
{A\approx-kT\left(\sum\limits _{i=1}^{n}\frac{N_{i}^{2}}{2V_{i}^{2}}I_{i}+\sum\limits _{i=1}^{n-1}\sum\limits _{\substack{j=i+1\\
j\leq n}
}^{i+3}\frac{N_{i}}{V_{i}}\frac{N_{j}}{V_{j}}I_{i,\, j}\right)}\\
{\mathrm{-}\sum\limits _{i=1}^{n}N_{i}kT\ln\frac{V_{i}}{N_{i}}+NkT(\ln\lambda^{3}-1)}\end{array}\label{GrindEQ__9_}\end{equation}

The pressure tensor {[}12, 13{]} and chemical potentials are obtained
from the following equations

\begin{eqnarray}
p_{xx,i} & = & -\frac{1}{L_{yi}L_{zi}}\left(\frac{\partial A}{\partial L_{xi}}\right)_{T,N}\notag\\
p_{yy,i} & = & p_{zz,i}=-\frac{1}{L_{xi}L_{yi}}\left(\frac{\partial A}{\partial L_{zi}}\right)_{T,N}\label{GrindEQ__10_}\\
\mu_{i} & = & \left(\frac{\partial A}{\partial N_{i}}\right)_{T,V,N_{j\neq i}}\notag\end{eqnarray}

If the system is in mechanical and chemical equilibrium, the $\mathit{xx}$\textit{\ } components of the pressure tensor and the chemical
potentials for each layer must be equal. From these equations, giving
as input the wall separation $\mathit{L}$ and the mean density $\rho^{\ast}=\rho\sigma^{3}$,
it is constructed a system of ($\mathit{n}-1$) non-linear equations
with ($\mathit{n}-1$) unknowns (layer densities and widths) to be
numerically solved. The low computational cost is taken for granted given that
the code is easily written and the calculations are carried out on
a Pentium 4 processor running at 2.66 GHz. At a temperature $\mathit{T}^{\ast}$=$\mathit{kT}$/$\varepsilon$=1,
we have explored the cases with $\mathit{L}$=10$\sigma$ and $\mathit{L}$=15$\sigma$,
at different mean densities. We have also compared the theoretical
results with experimental data coming from studies of nitrogen adsorption
in graphite slit pores at room temperature {[}19{]}.

\section{Monte Carlo simulation}

For numerical simulations, $\mathit{N}$ Lennard--Jones particles
are confined between hard walls separated at a distance $\mathit{L}$.
The unit cell is build up taken the walls to be of size $\mathit{L}_{\mathit{y}}$
and $\mathit{L}_{\mathit{z}}$\textit{\ }in the $\mathit{y}$ and
$\mathit{z}$ directions respectively, directions on which the periodical
boundary conditions are applied. The density profiles and pressures
were obtained taking average values in fluid slabs parallel to the
walls. The pressure tensor was used in the simple virial form, as
indicated in references {[}20, 21{]}.

With $\mathit{T}^{\ast}$=1, and for both slit pore widths $\mathit{L}$=10$\sigma$
and\textit{\ }$\mathit{L}$=15$\sigma$, the size of the unit cell
was set to $\mathit{L}_{\mathit{y}}$=$\mathit{L}_{\mathit{z}}$\textit{=}30$\sigma$,
taking the number of particles $\mathit{N}$ to correspond with the
mean density. The range of the Lennard--Jones interactions was considered
with a cutoff radius of 5$\sigma$.

\section{Results}

\begin{figure}
\begin{center}
\includegraphics[width=0.33\textwidth,angle=90]{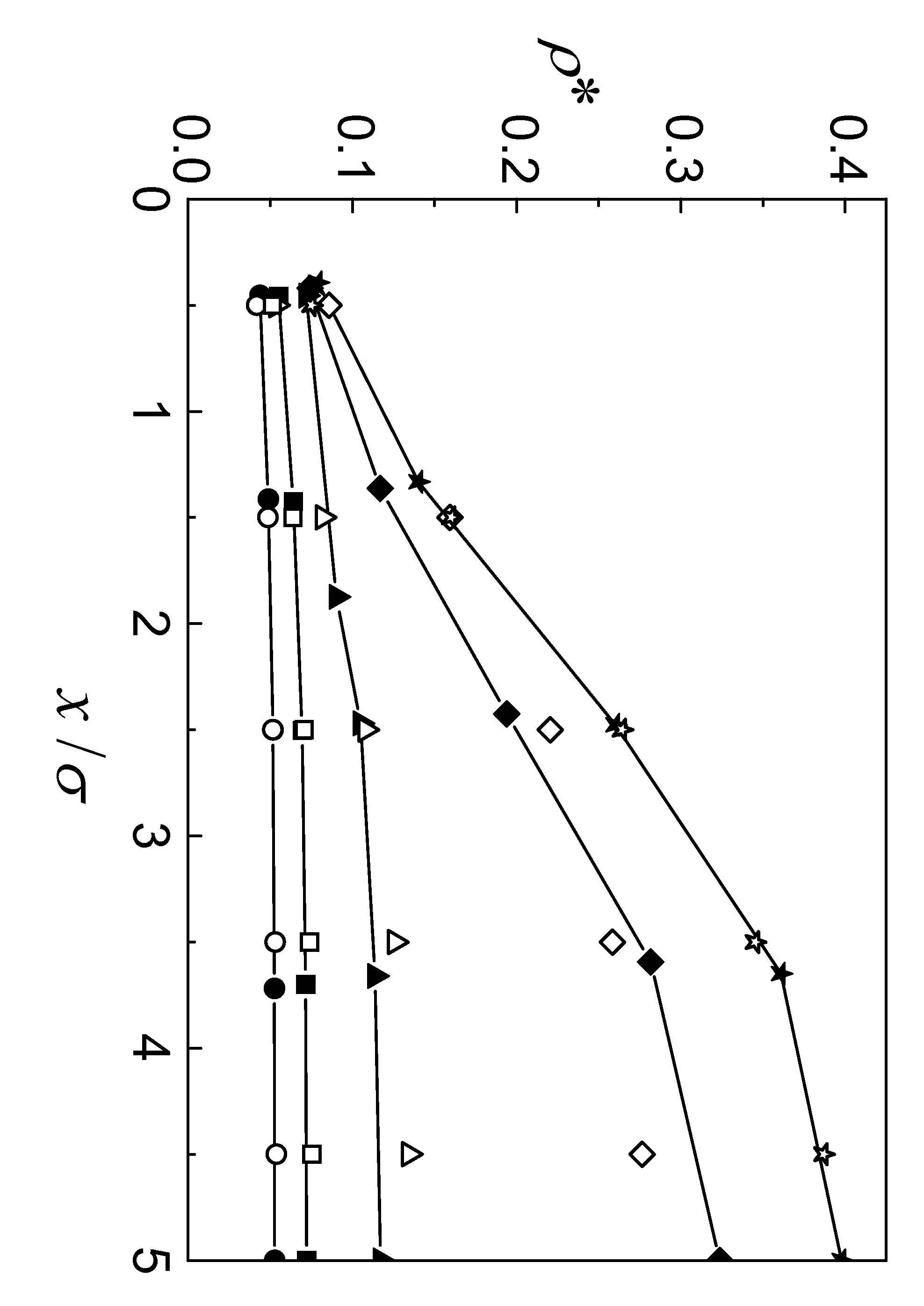} 
\end{center}
\caption{Density profiles for a $n$=9 layer model of a confined fluid in a
slit pore (solid symbols). The temperature is $\mathit{T}^{\ast}$=1.0 and the wall
separation is $\mathit{L}$=10$\sigma$, with mean densities $\rho^{\ast}$=1/20
(circles), 1/15 (squares), 1/10 (triangles), 1/5 (diamonds), and $\frac{1}{4}$
(stars). Open symbols represent the Monte Carlo simulations.}
\label{figure1} 
\end{figure}

\begin{figure}
\begin{center}
\includegraphics[width=0.33\textwidth,angle=90]{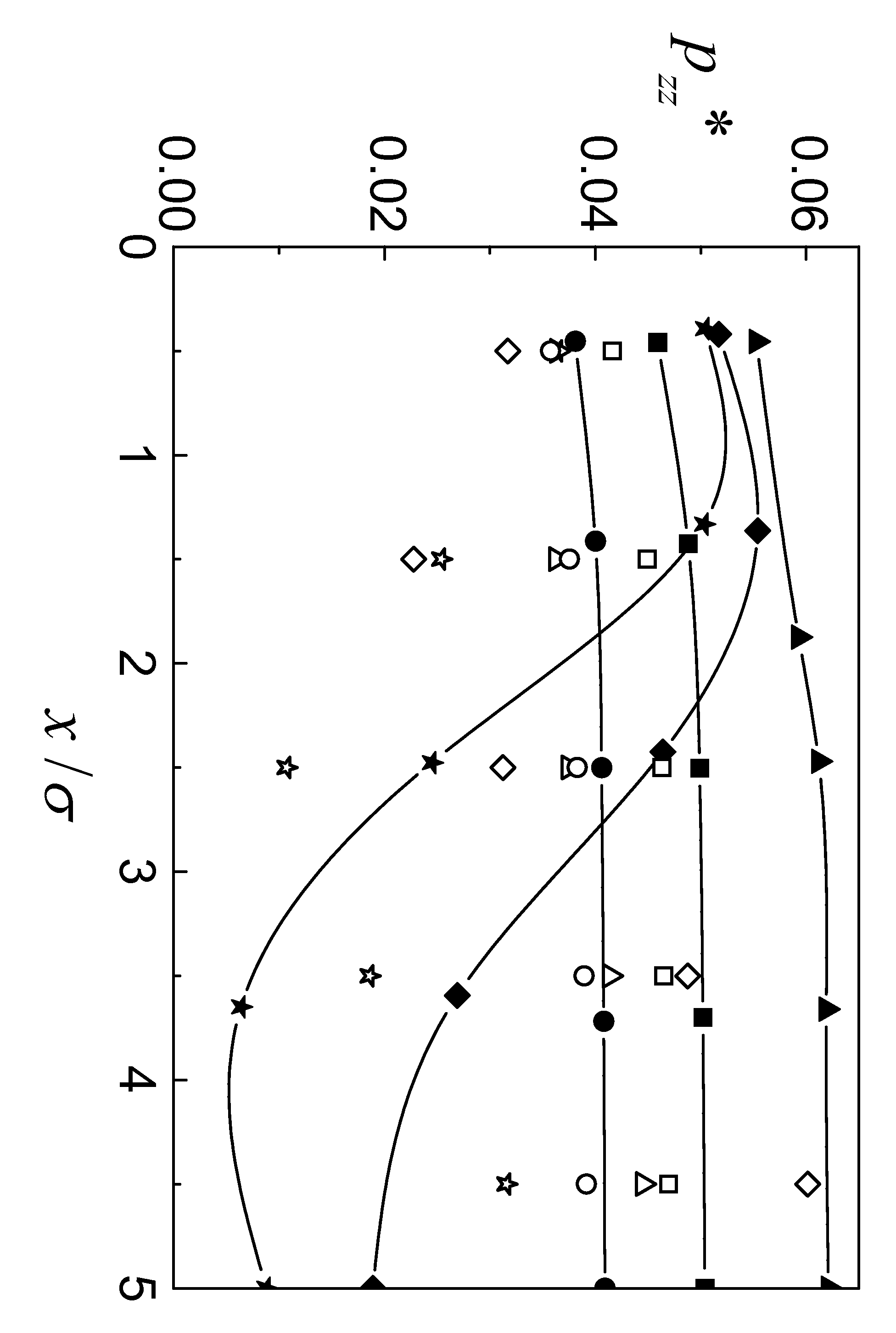} 
\end{center}
\caption{$\mathit{zz}$ pressure tensor component. Captions as in Fig. 1.}
\label{figure2} 
\end{figure}

In Figs. 1 and 2, the density profiles and $\mathit{zz}$ components
of the pressure tensor are shown for $\mathit{T}^{\ast}$=1 and\textit{\ }$\mathit{L}$=10$\sigma$.
The mean densities studied are $\rho^{\ast}$=1/20, 1/15, 1/10, 1/5,
and 1/4. The agreement of the theoretical density profiles with the
Monte Carlo simulations is very good. For the pressure there is a
rather good correspondence for low densities, up to $\rho^{\ast}$=1/10.
For the higher densities, differences appear, though the tendencies
are similar. The discrepancies come first from the low density approximations
done to get the Helmholtz energy. But, while in the simulation
slab particles fluctuate and at higher densities some clusterization
occurs, in the theory each layer is supposed to have a homogeneous
density which makes it hard for the theoretical pressures to follow those
obtained by simulation. For the density profiles, averaging the number
of particles in each slab evidently compensates the clusterization,
and the theory gives good results, at least for the rather low densities
studied. The same picture applies to the behavior of the system for
$\mathit{T}^{\ast}$=1 and\textit{\ }$\mathit{L}$=15$\sigma$, at
mean densities $\rho^{\ast}$= 1/10, and 1/5, represented in Figs.
3 and 4.

The results, as expected for hard repulsive walls, show a low density
region next to the walls and an increasing density profile, with a
maximum at the center of the slit pore. This behavior is also shown
with density functional theory {[}1{]} and in other Monte Carlo simulations
{[}2{]}.

\begin{figure}
\begin{center}
\includegraphics[width=0.33\textwidth,angle=90]{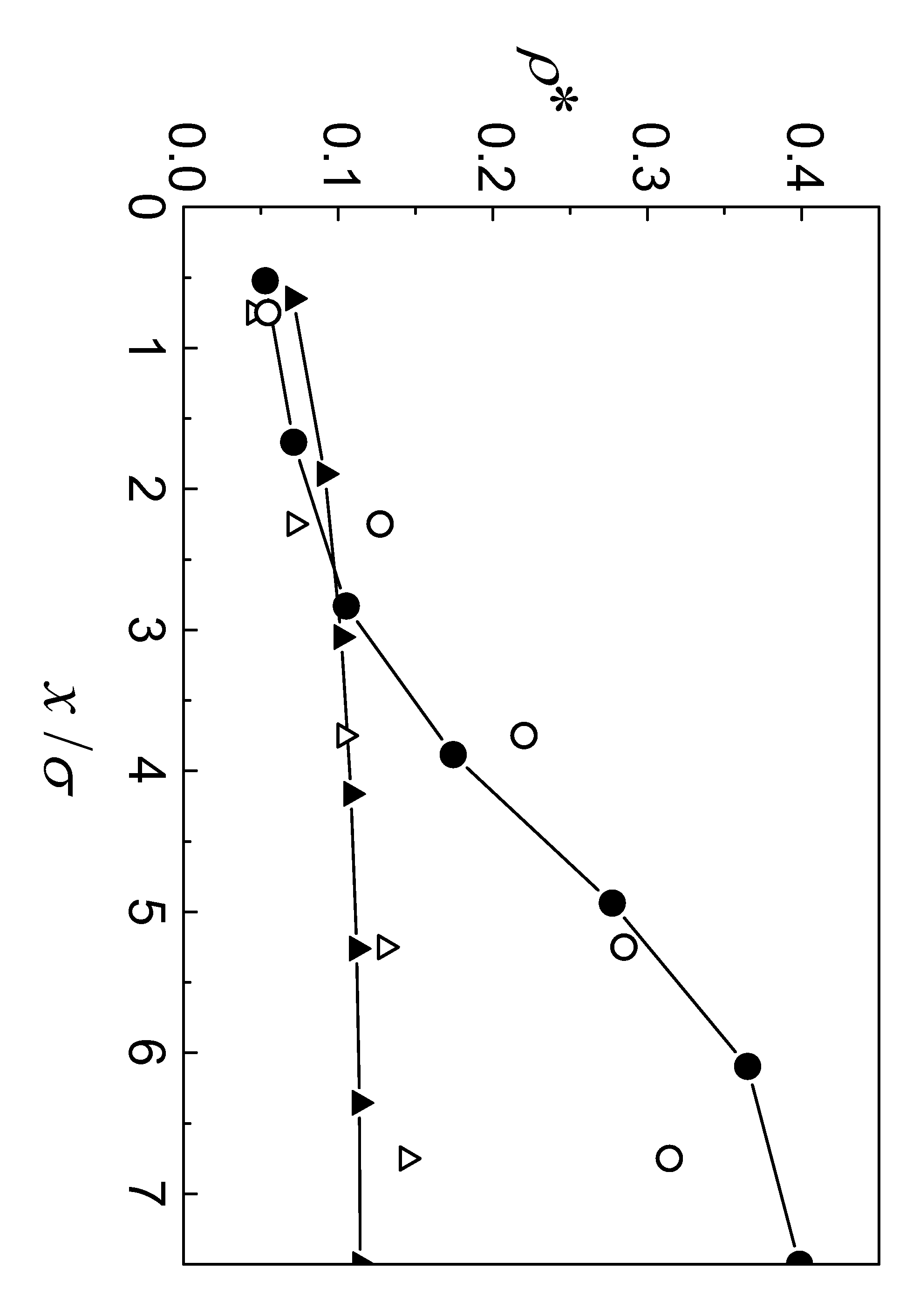} 
\end{center}
\caption{Density profiles for a $n$=13 layer model of a confined fluid in
a slit pore (solid symbols). The temperature is $\mathit{T}^{\ast}$=1.0 and the wall
separation is $\mathit{L}$=15$\sigma$, with mean densities $\rho^{\ast}$=1/10
(triangles), and 1/5 (circles). Open symbols represent the Monte Carlo
simulations.}
\label{figure3} 
\end{figure}

\begin{figure}
\begin{center}
\includegraphics[width=0.33\textwidth,angle=90]{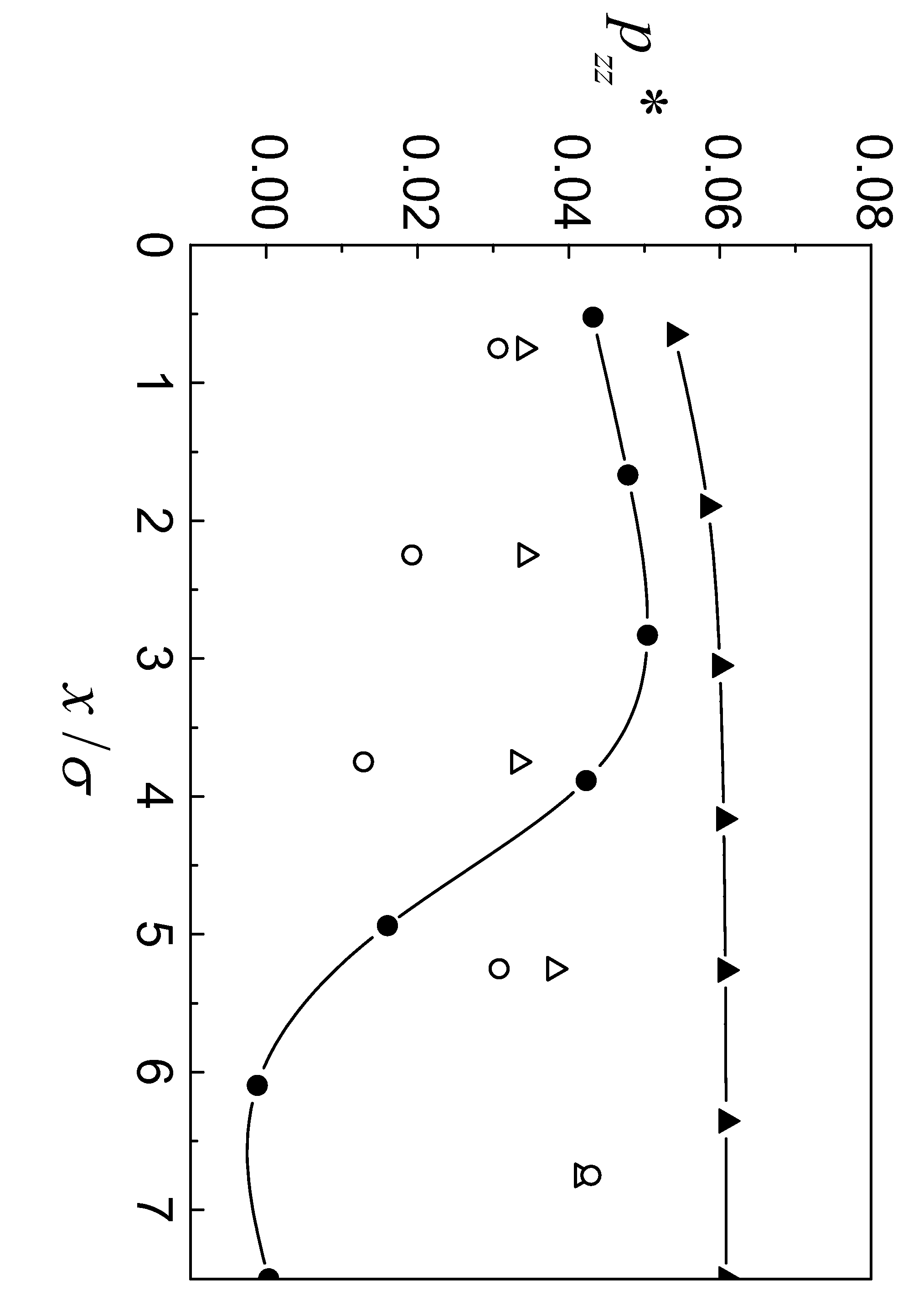} 
\end{center}
\caption{$\mathit{zz}$ pressure tensor component. Captions as in Fig. 3.}
\label{figure4} 
\end{figure}

Finally, the good agreement of the theory with the experiment can
be seen in the results shown in Fig. 5. In this figure, the excess
number of molecules per unit area of pore surface $\Gamma$ is plotted
in function of the external pressure, $\ $at $T^{\ast}$=3.18 and
$L$=4$\sigma$. These parameters approximate the experimental values
{[}19{]} $T$=303 K and $L$=1.45 nm, if $\varepsilon/k=95.2$ K and
$\sigma=3.75$ \AA{}\ are used to characterize the nitrogen. In this
case, due to the size of the sample, $n$=3 layers have been used
for calculation. $\Gamma$ is defined as

\begin{equation}
\Gamma=\frac{N-N_{g}}{S}=(\rho^{\ast}-\rho_{g}^{\ast})\frac{L}{\sigma^{3}}\label{GrindEQ__11_}\end{equation}
where $N_{g}$/$\rho_{g}^{\ast}$ is the number/density of particles
which would occupy the slit pore in the absence of the adsorption
forces. $\rho_{g}^{\ast}$ and the external pressure are determined
equating the chemical potential inside the slit pore (Eq. \ref{GrindEQ__10_})
to the chemical potential coming from the bulk van der Waals equation
at the same temperature. The theoretical results presented here are
similar to the numerical simulation results obtained by the same authors
who have done the experiment {[}19{]}. They assume that the differences
at higher pressures could be a consequence of the uncertainty in the
determination of the pore geometry.

\begin{figure}
\begin{center}
\includegraphics[width=0.47\textwidth]{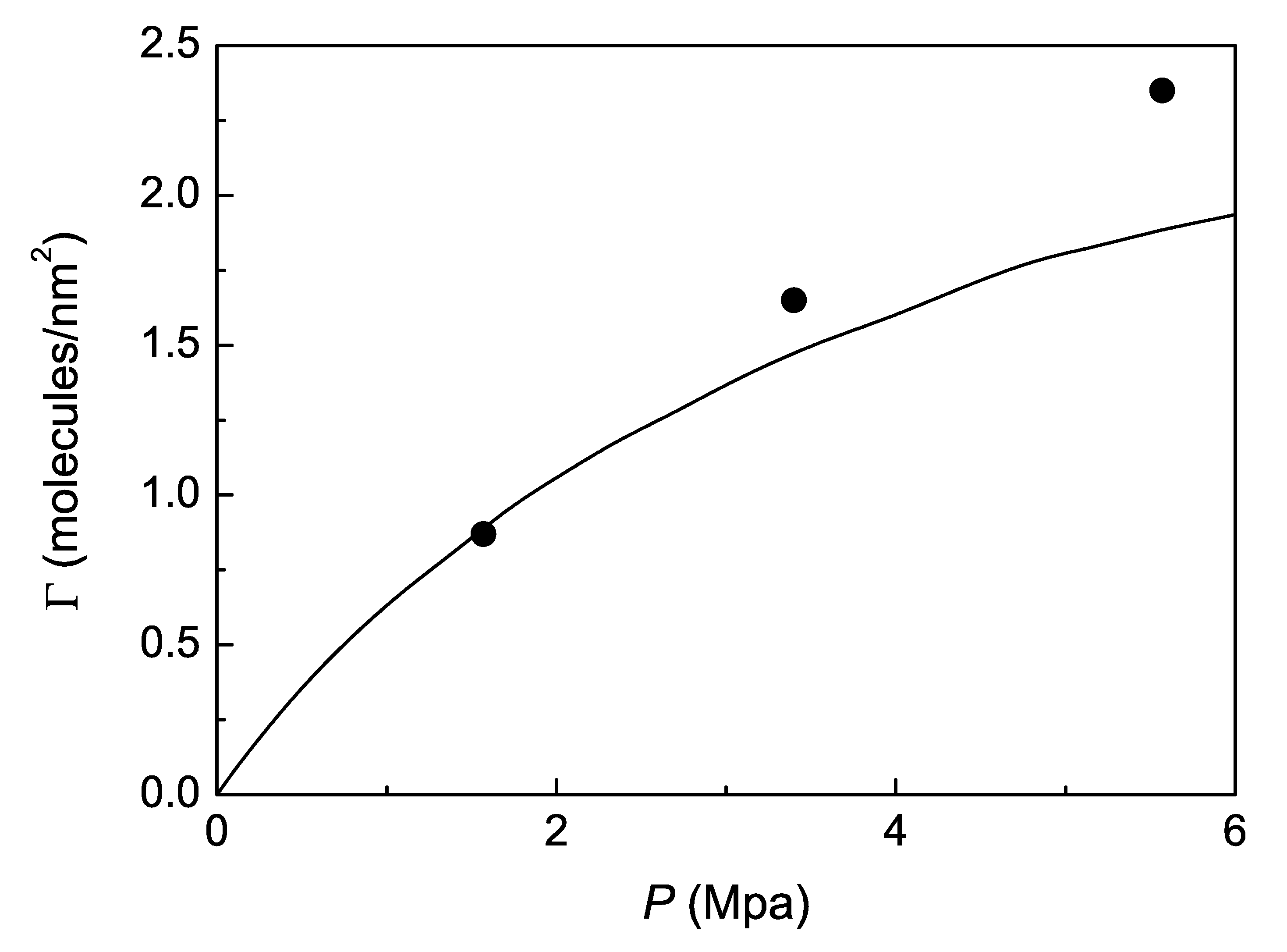} 
\end{center}
\caption{Excess number of molecules per unit area of pore surface $\Gamma$
as function of the external pressure. The full line represents the
experiment (digitalized from Ref. [19]), and the dots are our theoretical results.}
\label{figure5} 
\end{figure}

\section{Conclusions}

The application of a simple theory, with van der Waals-like approximations
to the Helmholtz energy, to a particular model of spatial distribution
makes it possible to obtain analytical expressions for the thermodynamic
quantities. The study of a confined fluid in a slit pore geometry
with a multilayer approximation produces good results when compared
with Monte Carlo simulations at low densities. The agreement with
a particular experiment on nitrogen confined in a graphite slit pore
is remarkable, even though an excess quantity is in study. It may
be concluded that the confinement reduces the importance that higher
virial contributions have on the equation of the state of the confined
fluid. Classical density functional theory {[}22{]} can also be applied
to study the slit pore geometry, with very good agreement with experiments
and simulations. Though the theoretical work developed in these
pages is not a competitor of density functional theory, it has the
advantages of having analytical expressions, and the possibility of
easily introducing two immiscible components: for instance one or
two layer lubricants wetting the walls and a gas or a liquid filling
the rest of layers forming the capillary volume. 

\begin{acknowledgements} This work was partially supported by Universidad
Nacional de La Plata and CICPBA. G. J. Z. is member of ``Carrera del
Investigador Cient\'{\i}fico'' CICPBA. \end{acknowledgements}

\section*{Appendix A}

Expressions of quantities used in Eqs. 5--8:

\bigskip
$\begin{array}{l}
{b=\frac{2}{3}\pi; B_{i}=\frac{\pi}{4}\frac{\sigma}{L_{xi}};A_{i}=a_{1}+\frac{a_{2}}{L_{xi}}+\frac{a_{3}}{L_{xi}^{3}}+\frac{a_{4}}{L_{xi}^{9}}}\\
\\{a_{1}=-\frac{16}{9}\pi;a_{2}=\frac{3}{2}\pi;a_{3}=-\frac{1}{3}\pi;a_{4}=\frac{1}{90}\pi} \hspace{0.5em}(A1)\end{array}$

\bigskip
A correction has been made to get good critical parameters for the
bulk ($\mathit{L\rightarrow\infty}$). For Argon $\mathit{a}_{1}$=
-5.7538 and $\mathit{b}$=1.3538, and for nitrogen $\mathit{a}_{1}$=
-1.5955 and \textit{b}=1.0349.

\bigskip
$\begin{array}{l}
{A_{i,\, i+1}=\frac{\pi}{90}\Bigl[ -\frac{1}{L_{xi}^{9}}-\frac{1}{L_{xi}L_{xi+1}^{8}}+\frac{1}{L_{xi}(L_{xi}+L_{xi+1})^{8}}\Bigr] }\\
\hspace{3em}-\frac{\pi}{3}\Bigl[ -\frac{1}{L_{xi}^{3}}-\frac{1}{L_{xi}L_{xi+1}^{2}}+\frac{1}{L_{xi}(L_{xi}+L_{xi+1})^{2}}\Bigr] \\
\hspace{3em}-\frac{3}{2}\frac{\pi}{L_{xi}}\hspace{13em}(A2)\end{array}$

\bigskip
$\begin{array}{l}
A_{i,\, i+2}=\frac{\pi}{90}\Bigl[\frac{1}{L_{xi+1}^{8}}-\frac{1}{(L_{xi}+L_{xi+1})^{8}}\\
\hspace{3em}-\frac{1}{(L_{xi+1}+L_{xi+2})^{8}}+\frac{1}{(L_{xi}+L_{xi+1}+L_{xi+2})^{8}}\Bigr] \frac{1}{L_{xi}}\\
\hspace{3em}-{\frac{\pi}{3}\Bigl[\frac{1}{L_{xi+1}^{2}}-\frac{1}{(L_{xi}+L_{xi+1})^{2}}-\frac{1}{(L_{xi+1}+L_{xi+2})^{2}}}\\
\hspace{3em}{+\frac{1}{(L_{xi}+L_{xi+1}+L_{xi+2})^{2}}\Bigr]\,\frac{1}{L_{xi}}}\hspace{5em}(A3)\end{array}$ 

\bigskip
$\begin{array}{l}
{A_{i,\, i+3}=\frac{\pi}{90}\Bigl[\frac{1}{(L_{xi+1}+L_{xi+2})^{8}}}\\
\hspace{3em}-\frac{1}{(L_{xi}+L_{xi+1}+L_{xi+2})^{8}}\\
\hspace{3em}-\frac{1}{(L_{xi+1}+L_{xi+2}+L_{xi+3})^{8}}\\
\hspace{3em}+{\frac{1}{(L_{xi}+L_{xi+1}+L_{xi+2}+L_{xi+3})^{8}}\Bigr]\frac{1}{L_{xi}}}\\
\hspace{3em}-{\frac{\pi}{3}\Bigl[\frac{1}{(L_{xi+1}+L_{xi+2})^{2}}-\frac{1}{(L_{xi}+L_{xi+1}+L_{xi+2})^{2}}}\\
\hspace{3em}{-\frac{1}{(L_{xi+1}+L_{xi+2}+L_{xi+3})^{2}}}\\
\hspace{3em}+{\frac{1}{(L_{xi}+L_{xi+1}+L_{xi+2}+L_{xi+3})^{2}}\Bigr]\,\frac{1}{L_{xi}}}\hspace{2em}(A4)\end{array}$

\end{document}